%% file: approximate_matrix_multiplication_using_UEP_v1.0.tex
\documentclass[conference]{IEEEtran}
\IEEEoverridecommandlockouts
\usepackage[noadjust]{cite}
\usepackage{amsmath,amssymb,amsfonts}
\usepackage{dsfont}
\usepackage{graphicx}
\usepackage{textcomp}
\usepackage{setspace}
\usepackage{xcolor}
\usepackage{multirow}
\usepackage{amsthm}
\usepackage[noend]{algpseudocode}
\usepackage{subcaption}
\usepackage{mathcomSTEv4}

\usepackage[textfont=normalfont,singlelinecheck=off,justification=raggedright,font=footnotesize]{caption}

\usepackage[margin=30pt,textfont=normalfont,singlelinecheck=off,justification=raggedright,font=footnotesize]{subcaption}
\usepackage{dsfont}

\usepackage{environ}         
\usepackage{etoolbox}        
\usepackage{graphicx}        

\usepackage{anyfontsize}         

\usepackage{xcolor}
\usepackage{multicol}
\usepackage[utf8]{inputenc}

\usepackage[hidelinks]{hyperref}
\usepackage{booktabs}
\usepackage{bbm}
\usepackage{authblk}

\usepackage{tikz,pgfplots}
\usetikzlibrary{arrows}

\newtheorem{theorem}{Theorem}

\usepackage{caption,booktabs}

\makeatletter
\def\endthebibliography{%
	\def\@noitemerr{\@latex@warning{Empty `thebibliography' environment}}%
	\endlist
}

\makeatother
\begin{document}

\title{
%
%
Straggler Mitigation through Unequal Error Protection for Distributed Matrix Multiplication
}
\author[*]{Busra Tegin}
\author[$\dagger$]{Eduin E. Hernandez}
\author[$\dagger$]{Stefano Rini}
\author[*]{Tolga M. Duman}
\affil[*]{Department of Electrical and Electronics Engineering \authorcr 
Bilkent University, Turkey \authorcr
Email: \{btegin, duman\}@ee.bilkent.edu.tr }
\affil[$\dagger$]{Department of Electrical and Computer Engineering \authorcr 
National Yang Ming Chiao Tung University, Taiwan \authorcr
Email: \{eduin.ee08, stefano.rini\}@nycu.edu.tw }

\renewcommand\Authands{ and }

\maketitle

\begin{abstract}
Large-scale machine learning and data mining methods routinely distribute computations across multiple agents to parallelize processing. 
The time required for computation at the  agents is affected by the availability of local resources giving rise to the ``straggler problem'' in which the computation results are held back by unresponsive agents.
For this problem, linear coding of the matrix sub-blocks can be used to introduce resilience toward straggling. 
The Parameter Server (PS) utilizes a channel code and distributes the matrices to the workers for multiplication. 
It then produces an approximation to the desired matrix multiplication using the results of the computations received at a given deadline. 
%
%
In this paper, we propose to employ Unequal Error Protection (UEP) codes to alleviate the straggler problem.
The resiliency level of each sub-block is chosen according to its norm as blocks with larger norms have higher effects on the result of the matrix multiplication. 
%
%
%
%
%
We validate the effectiveness of our scheme both theoretically and through numerical evaluations.
We derive a theoretical characterization of the performance of UEP using random linear codes, and compare it the case of equal error protection. 
We also apply the proposed coding strategy to the computation of the back-propagation step in the training of a Deep Neural Network (DNN), for which we investigate the fundamental trade-off between precision and the time required for the computations. 
%
%
%
%
\end{abstract}


\section{Introduction}
Distributed computing clusters are fundamental in many domains, such as machine learning, data-mining, and high-precision numerical simulations as they allow parallelization of the computational tasks \cite{baldini2017serverless}.
The widespread reliance on distributed computation clusters presents several opportunities over traditional computing paradigms, but also offer a new set of challenges.
Among the most well-recognized issues is that of the stochasticity in the time required for the computation. 
This gives rise to the phenomenon of ``stragglers'', that is, agents with large response times which delay computation.  
As a remedy, channel coding can be applied to reduce the delays in computation due to stragglers \cite{reisizadeh2019coded}.
In this paper, we propose a novel scheme for distributed computation with stragglers which makes use of the  variations in the magnitude of the matrix entries which naturally occur in many applications, such as back-propagation in Deep Neural Network (DNN) training.
%
We first identify the matrix sub-products which are expected to have the largest norms and use Unequal Error Protection (UEP) to provide a level of resiliency against stragglers.
The proposed solution offers an improved resilience by providing an improved approximate reconstruction of the matrix product by a given computation deadline.
%
%
%

\subsection{Literature Review}
As matrix multiplication is a fundamental algebraic operation, distributed approximate matrix multiplication has been investigated in many contexts. 
In the big-data paradigm, computation and storage are distributed, hence computer processing architectures can be devised for efficiently performing this operation \cite{choi1994pumma,van1997summa}.
In a cloud-computing setting, distributed matrix computation is investigated in \cite{gupta2018oversketch,kim2019mpec}.
Distributed matrix computation for DNN training through back-propagation which involves multiplication of large matrices is studied in
%
\cite{plancher2019application,son2018distributed}.
More recently, the problem of ``stragglers'' has been recognized as an important issue. 
In many distributed computation platforms such as Amazon Web Services Lambda and Google Cloud Functions, distributed computation can be held back by a set of workers which take much longer than the median job time \cite{dean2013tail}. 
Such  random delays  decrease the overall computational efficiency of the system.
To mitigate the effect of straggles, coding for matrix multiplication can be applied \cite{lee2017speeding}.
Since its inception in \cite{lee2017speeding}, this line of research received significant attention in the literature.
In \cite{wang2015using}, the authors use  the theory of extreme order statistics to analyze how task replication reduces latency. 
In \cite{dutta2016short}, the authors  introduce redundant computations in a coding theory inspired fashion, for computing linear transforms of long vectors. 
Product codes for distributed matrix multiplication are studied in \cite{baharav2018straggler}.
A new class of codes, called polynomial codes, is proposed in \cite{yu2017polynomial}, and their optimality is argued for the straggler problem.
While the above literature focuses on minimizing the time for completing a computation task, one can also consider approximate computation.
Along these lines, in \cite{gupta2018oversketch}, the authors propose OverSketch, an algorithm that uses matrix sketching to approximate matrix multiplication.

\subsection{Contribution}

In this paper, we investigate the trade-off between accuracy and delay in distributed approximate matrix multiplication with stragglers. %
Multiplication of large sparse matrices is an important problem in implementing machine learning algorithms. 
Since for typical machine learning problems only approximate matrix multiplication results are sufficient, we consider a distributed matrix multiplication scheme in which the sub-blocks of the matrices being multiplied are encoded using UEP codes and they are distributed across different workers.
The workers respond with the results of the products (of coded sub-blocks), effectively resulting in UEP coded sub-products of the two matrices. 
The main goal is to produce an approximation of the product of two matrices as quickly as possible; with a more and more accurate approximation with more and more workers responding, i.e., a progressively improving matrix approximation in time, exploiting the UEP code constraints.

Our main contribution is the proposal of employing UEP codes to improve the quality of the approximation of matrix multiplications by exploiting the variations in the matrix entries' magnitudes. 
By carefully matching the matrix sub-products' norms with the level of unequal error protection, we demonstrate that significant improvements can be attained over the case of equal error protection or uncoded computation. 

The paper is organized as follows. In Section \ref{sec:System Model}, we formulate the distributed approximate matrix multiplication problem by considering the case in which the multiplication is broken down in a set of row-times-column products distributed to workers which can perform sub-matrix multiplications with a random computation time. 
In Section \ref{sec:Approximate Matrix Multiplication with UEP Codes}, we present our  proposed scheme in which UEP codes are used to code the row and column terms of the block matrix multiplication. 
In particular, we leverage the construction in \cite{vukobratovic2012unequal} to offer more protection  to the  sub-products with larger norms and reduce the effects of the randomness in the service time.
In Section \ref{sec:Theoretical Analysis}, we provide a theoretical evaluation of the expected error in the matrix approximation as a function of the service time distribution.
In Section \ref{sec:Numerical Examples}, we present the results of a DNN training when the back-propagation step is distributed among workers as in the proposed scheme.
This example is particularly relevant since the matrices of the different DNN layers have different levels of sparsity, and thus, they result in highly varied matrices to be multiplied. 
Furthermore, we illustrate the performance of the proposed strategy using simple matrix models, and compare it with those of Maximum Distance Separable (MDS) codes.
The paper is concluded in Section \ref{sec: conclusion}.

\smallskip

\noindent
{\bf Notation:} Matrices are denoted with bold capital Roman letters, e.g. $\Av$, column vectors with bold lower-case roman letters, e.g. $\vv$. 
The Frobenius  norm of the matrix $\Av$ is shown as $\| \Av \|_F$. 
%
The set of integers $\{1,\ldots,N\} \subset \Nbb$ is denoted as $[N]$.
Given two matrices $\Av_1$ and $\Av_2$ with the same number of rows, we depicted their column-wise concatenation as $\Av=[\Av_1 \: , \: \Av_2]$. Similarly, given $\Av_1$ and $\Av_2$ with the same number of columns, their row-wise concatenation is represented as $\Av=[\Av_1 \: ; \: \Av_2]$ which can also be equivalently expressed as $\Av=[\Av_1^{\intercal} \: , \: \Av_2^{\intercal}]^{\intercal}$.
Capital Roman letters are used for scalars. 
Finally, $\mathcal{N}(\mu, \sigma^2)$ indicates the Gaussian distribution with mean $\mu$ and variance $\sigma^2$.
%
%

\section{System Model}
\label{sec:System Model}
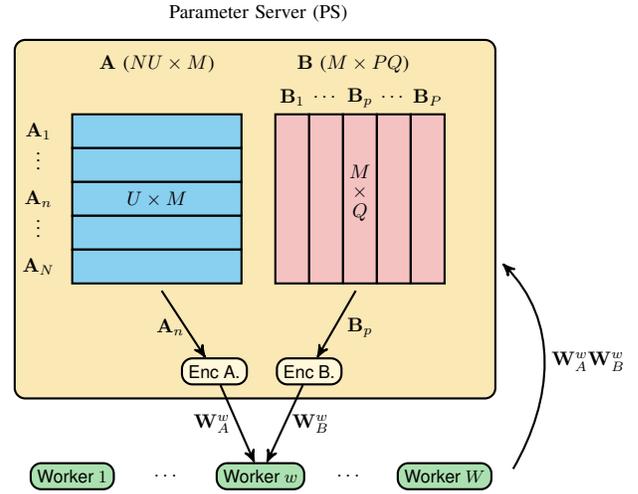
\begin{figure}[t]
	\centering
    \input{tikz_files/system_model}
	\caption{System model in Section \ref{sec:System Model}.}  
	\label{systemmodel}
	\vspace{-0.5cm}
\end{figure}

We consider the scenario in Fig. \ref{systemmodel} in which a PS wishes to compute the matrix product 
$\Cv=\Av \Bv$ by distributing the row-times-column matrix  multiplications  among $W$ workers. 
%
%
Each worker receives a fat and a tall matrix, computes their product and returns it to the PS. 
We assume that there is no communication cost between the workers and the PS, and vice versa. 
The time required for computation is a random variable.
At a given time, the PS produces an approximation $\Chv$ of the matrix $\Cv$ by using the received sub-products from the workers.

Let us consider the problem setting in more detail. 
Consider the matrices $\Av$ and $\Bv$ with elements  from a finite field $\mathbb{F}$,  and with dimensions  $N U \times M$ and $M \times P Q$, respectively. 
The aim of the PS is to produce  an approximate expression for the matrix  multiplication $\Cv=\Av \Bv$ as $\Chv$ with respect to the loss\footnote{In the following, we only consider the case of a Frobenius norm: the case of a more general loss is not discussed here for brevity.}
\ea{
\Lcal (\Cv,\Chv) = \|\Cv-\Chv\|_F^2. \tag{1}
\label{eq:loss}
}

To accomplish this, the PS  divides the matrix products into sub-products and distributes them across a set of workers. 
Specifically, as in \cite{lee2017high}, we partition $\Av$/$\Bv$ into row/column sub-blocks $\Av_n$/$\Bv_p$ as 
\begin{align} \tag{2} \label{eq:sub-blocks}
 \Av & = [\Av_1 \:;\: \cdots  \Av_n \cdots\:;\: \Av_N], \\
\Bv & = [\Bv_1  \:,\: \cdots  \Bv_p  \cdots \:,\: \Bv_P], \nonumber
\end{align}
where $\Av_n  \in \mathbb{F}^{U  \times M}$ and $\Bv_p  \in \mathbb{F}^{M\times Q}$ for $ n \in [N]$, $p\in [P]$. 
The $n\times p$-th sub-product is $\Cv_{np} \in \mathbb{F}^{U  \times Q}$. Clearly, $NP$ such matrix multiplications are needed to produce $\Cv$. 
In general, not all the sub-blocks have the same impact on the final  matrix multiplication result, as some sub-blocks may have larger Frobenius norms compared to the others.
This motivates the use of UEP codes to better protect the more impactful sub-products when distributing the computation to the workers. 
Accordingly, we consider that the PS sends to each worker $w$ the matrices $\Wv_{A}^w$ and $\Wv_{B}^w$  obtained as 
\ea{ \tag{3} \label{eq:encode}
\Wv_{A}^w & = f_{ {\rm enc} - A} \lb \Av_1, \cdots , \Av_N \rb, \\
\Wv_{B}^w & =  f_{{\rm enc} - B} \lb \Bv_1, \cdots , \Bv_P \rb, \nonumber
}
and sets a time deadline $T_{\max}$ by which it expects the matrix product $\Wv_{A}^w \Wv_{B}^w$ to be returned.
At time $T_{\max}$, the PS produces the matrix approximation
\ea{
\Chv = f_{ {\rm dec}-C} \lb \Wcal(T_{\max})\rb, \tag{4}
\label{eq:Ch}
}
where $\Wcal(T_{\max})$ is the set of matrix products received up to time $T_{\max}$.
Using  $\Chv$, the loss in \eqref{eq:loss} can be evaluated. %
Let us denote it as $\Lcal(T_{\max})$.
The  random set $\Wcal(T_{\max})$ is obtained as follows:
We assume that the computation time of the worker $w$ is equal to the random variable $T_w$ which is identical and independently distributed (i.i.d.) with a cumulative distribution function (CDF) $F(t)$. 
Accordingly, the set $\Wcal(T_{\max})$ contains the products $\Wv_{A}^w \Wv_{B}^w$ for which $T_w<T_{\max}$.

\smallskip

The problem we consider next is to design the functions $f_{ {\rm enc} - A}$,$f_{ {\rm enc} - B}$ and $f_{ {\rm dec} - C}$ such that the loss in \eqref{eq:loss} is minimized over some dataset of matrix multiplications $\Dcal(\{ \Av,\Bv\})$.

\section{Approximate Matrix Multiplication\\ Using UEP Codes}
\label{sec:Approximate Matrix Multiplication with UEP Codes}
In this section, we describe a distributed coded approximate matrix multiplication scheme which aims to provide better protection for the matrix sub-products $\Av_n\Bv_p$ with  larger norms, and thus produce a better approximation to the product of two matrices within the prescribed time.
The coding scheme is parametrized in such a way that matches the distribution of the matrices in the dataset.

\subsection{Importance Level of a Sub-block}

Let us begin by classifying the matrix sub-block in \eqref{eq:sub-blocks} according to their norms. For instance,  we may select three different levels for each sub-block, e.g., \textit{high}, \textit{medium}, and \textit{low} to classify the norms of $\Av_n$ and $\Bv_p$.
Let us refer to these levels as \emph{importance levels}, and assume that there are $S$ such levels.
Given a matrix $\Av$/$\Bv$, we have $n_A^s$/$n_B^s$ blocks with level of importance $s \in [S]$ (let the importance be decreasing in $s$).
Clearly, $N =\sum_{s \in [S]} n_A^s$, and $P =\sum_{s \in [S]} n_B^s$.

By construction, any sub-product $\Cv_{np}$ is obtained as the multiplication of sub-blocks in two classes: accordingly $\Cv_{np}$ has $L$ possible importance levels with $L = S(S+1)/2$.
For instance, in the examples of three importance levels for $\Av_n$ and $\Bv_p$,  $\Cv_{np}$ can have importance  \textit{high $\times$ high}, \textit{high $\times$ medium}, \textit{high $\times$ low}, \textit{medium $\times$ medium}, etc.
%
%
%

From a high level-perspective, one would want the PS to be able to more quickly recover those products corresponding to the importance level high $\times$ high,  while the importance level low $\times$ low is not particularly urgent. 
We can obtain this desired behavior by employing UEP codes.
%
%
%

\subsection{UEP Coded Matrix Multiplication}
While there are different ways of UEP coding, we focus on the case in which the encoding functions in \eqref{eq:encode} are the UEP codes described in \cite{vukobratovic2012unequal}. 
Specifically, we consider the use of two different UEP schemes called Non-Overlapping Windows (NOW) UEP and Expanding Window (EW) UEP strategies based on Random Linear Codes (RLC).
Let us briefly introduce these codes next.

The NOW-UEP coding strategy simply divides the packets into ``windows'' based on their importance, and applies an RLC for each type independently and separately. 
The encoding is performed firstly by selecting a window using a window selection polynomial function $\Gamma(\xi)= \sum_{i \in [L]} \Gamma_i \xi^i$, where $L$ is the number of block types, and $\Gamma_i$ is the window selection probability for the $i$-th type. 
The encoded matrices are generated only from the matrices of the selected type. 
The PS selects importance levels for both $\Av$ and $\Bv$ independently using predetermined window selection distributions, and encodes the corresponding rows and columns of $\Av$ and $\Bv$ as
\begin{equation} \tag{5}
\begin{aligned}
\Wv_{A}^w &= \sum_{i } \alpha_{i}^w \Av_{\pi_A^w(i)},\\
\Wv_{B}^w &= \sum_{j} \beta_{j}^w \Bv_{\pi_B^w(j)},
\end{aligned}
\end{equation}
for the $w$-th worker, where $ \alpha_{i}^w$ and $ \beta_{j}^w$ are randomly
selected elements from the given finite field, and
${\pi_A^w(i)}$/${\pi_B^w(i)}$ are the row/column indices of $\Av$/$\Bv$ at the corresponding levels, respectively. 


The EW-UEP coding strategy also uses a probabilistic window selection polynomial $\Gamma(\xi)$ for row and column class selection of $\Av$ and $\Bv$, respectively; however, the window definition is different from that of the NOW-UEP strategy. 
The EW-UEP constructs the $i$-th window by including all the packets whose importance levels are $i$ or higher than $i$. 
For instance, let us assume that the third importance level is selected according to the window selection distribution. This strategy dictates inclusion of all the source messages from the first, second, and third importance levels in the codeword.
Thus, the EW strategy includes the most important matrices in the encoding process regardless of the importance level of the selected window to provide a better protection than others.

In both cases, the decoding function simply places $\Chv_{np}=\Cv_{np}$ when it is possible to obtain it from $\Wcal(T_{\max})$, and sets $\Chv_{np}$ to the all-zero matrix otherwise.

\begin{table}[t]
\vspace{0.5cm}
    \centering
        \caption{
        A summary of the quantities in Section \ref{sec:System Model} and Section  \ref{sec:Approximate Matrix Multiplication with UEP Codes} and their respective indexes. }
    \label{tab:my_label}
    {
\begin{tabular}{|c|c|c|c|}
\hline
Matrix      & Size           & Constant & Value \\ \hline
$\mathbf A$ & $NU \times M$  & \# Workers     & $W$     \\ \hline
$\mathbf B$ & $M \times PQ$  & \#  Importance levels ($\mathbf{A}/ \mathbf{B}$)     & $S$     \\ \hline
$\mathbf C$ & $NU \times PQ$  & \#  Importance levels ($\mathbf C$)     & $L$     \\ \hline
$\Av_n$ & $U \times M$ & \#  Row blocks     & $N$  \\ \hline
$\Bv_p$ & $M \times Q$ & \#  Column  blocks & $P$ \\ \hline
$\Cv_{np}$ & $U \times Q$ & Computation deadline & $T_{\max}$ \\ \hline
\end{tabular}
}
\vspace{-0.5cm}
\end{table}

\section{Theoretical Analysis}
\label{sec:Theoretical Analysis}

In \cite{vukobratovic2012unequal}, the authors give the corresponding decoding probabilities of NOW and EW-UEP strategies for each importance level as a function of the number of received packets in each class. 
In this section, we assume for simplicity that the entries of the matrices are zero mean and with variance  $\sigma_{A_n}^2$ and $\sigma_{B_p}^2$ in each sub-block, and they are uncorrelated, so that $\mathbb{E} \Big[ \|\Cv_{np} \|_F^2 \Big]  = MUQ \sigma_{A_n}^2 \sigma_{B_p}^2$.
Let us denote the number of encoded matrix products received at time $t$ by $N(t)$, then the probability of receiving $w$ packets from $W$ workers at time $t$ is $P_{N(t)}(w)$, which is simply calculated as
\begin{equation} \tag{6}
P_{N(t)}(w) = {W \choose w} (1-F(t))^{W-w} \left(F(t)\right)^{w}.
\end{equation}

%

From  \cite[Eq. 5]{vukobratovic2012unequal}, we obtain a bound (which is achievable with large field sizes) on the decoding probabilities of NOW-UEP strategy for each importance level as a function of received matrices as 
\begin{equation} \tag{7} \label{dec_prob}
P_{d,l}(N)\leq \sum_{\substack{(n_{1},n_{2},\ldots,n_{L}) \\
		\sum_{i \in [L]} n_{i}=N }}
	P_{\Gamma(\xi),N}(\mathbf{n}) \: \onev(n_{l}\geq k_{l}),
\end{equation}
where $\mathbf{n} = [n_{1},n_{2},\ldots,n_{L}]$, $k_{l}$ is the number of packets in class $l$ and $\onev(\cdot)$ is the indicator function, and 
\begin{equation} \tag{8}
P_{\Gamma(\xi),N}({\bf n})=\frac{N!}{n_{1}!n_{2}!\ldots n_{L}!} \Gamma_{1}^{n_{1}}\Gamma_{2}^{n_{2}}\ldots\Gamma_{L}^{n_{L}}.
\end{equation}
As an example, with three classes, $W=40$ workers, and window selection probabilities $(0.35,0.35,0.3)$, the decoding probabilities of each class are as depicted in Fig. \ref{thm:NOW-UEP}, which clearly illustrates that the most important class is protected better. 

We can bound the performance of the coded matrix multiplication scheme in Section \ref{sec:Approximate Matrix Multiplication with UEP Codes} as follows.  
%
\begin{theorem}{\bf NOW-UEP Loss:}
\label{thm:NOW-UEP}
Consider the loss minimization problem in Section \ref{sec:System Model} for the case in which the set of matrix products $\Dcal(\{\Av,\Bv\})$ is the set of matrices with i.i.d. entries with variance $\sigma_{l,A}^2$ and  $\sigma_{l,B}^2$ for the $l$-th class of the final product $\Cv$, respectively, for $\Av$ and $\Bv$. 
The expected loss of approximate matrix multiplication with the NOW-UEP strategy described in Section \ref{sec:Approximate Matrix Multiplication with UEP Codes} is 
\begin{equation} \label{mse_t} \tag{9}
\Ebb\lsb \Lcal (T_{\max})\rsb =  \sum_{w \in [W]}  P_{N(T_{\max})}(w) \mathbb{E} [\|\Cv- \hat{\Cv}\|_F^2 | N(t) = w ],
\end{equation}
where 
\begin{align} 
 \mathbb{E} [\|\Cv-& \hat{\Cv}\|_F^2 | N(t)=w ] \nonumber  \\
 &= \sum_{l \in [L]} k_l \left(1-P_{d,l}(w)\right)
MUQ \sigma_{l,A}^2 \sigma_{l,B}^2, \tag{10}
\end{align}
	where $k_l$ is the number of blocks in the $l$-th importance level of $\Cv$, and the expectation in \eqref{mse_t} is taken over the random entries of $\Av$, $\Bv$.
\end{theorem}
Note that since \eqref{dec_prob} is an upper bound on the correct recovery probability, applying it to \eqref{mse_t} results in a (lower) bound on the expected loss, however, this bound is tight as the field size tends to infinity, i.e., the lower bound on the loss is asymptotically achievable. 
%
The analog of Theorem \ref{thm:NOW-UEP} for the EW-UEP is obtained from the results in \cite[Eq. 4-9]{ vukobratovic2012unequal} and is not presented here for brevity.
Note that, in  Th. \ref{thm:NOW-UEP}, there exists a ``matching'' between the probabilistic structure of the matrices to be multiplied and their row/column block partitioning.
In reality, one would not observe such a neat organization of the matrix values.
Instead one would have to fit the row/column weight distribution in the data to design the UEP code resulting in the minimal loss.

\section{Numerical Evaluations}
\label{sec:Numerical Examples}

We now provide examples with the approximate matrix multiplication scheme using UEP codes proposed in Section \ref{sec:Approximate Matrix Multiplication with UEP Codes}.
For service time of workers, we use the exponential latency model, i.e., the completion time of each worker for a given task is exponentially distributed with parameter $\lambda$. 

\subsection{Matrix Approximation Using UEP}
\label{subsec:Matrix Approx UEP}

\begin{figure}[t]
	\centering
	\vspace{-3cm}
    \input{tikz_files/NOW_prob_40workers}
	\vspace{-0.4cm}
	\caption{Decoding probabilities of NOW-UEP strategy with three classes, and $W=40$ workers.}  \label{NOW_prob}
	\vspace{-0.5cm}
\end{figure}
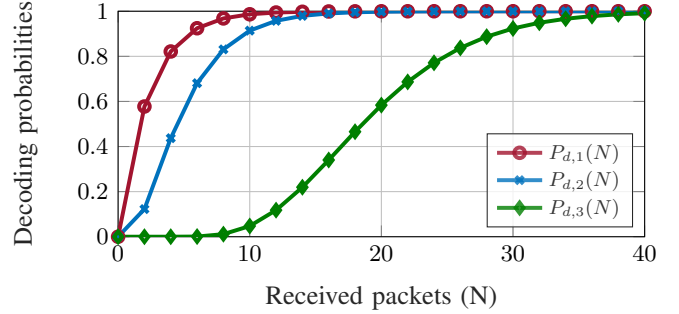

\begin{figure}[t]
	\centering
    \input{tikz_files/three_classes_UEP_t_40workers}
	\vspace{-0.2cm}
	\caption{Normalized loss of the estimator using UEP codes with three classes with exponential latency model.}  \label{three_cls_mse}
	\vspace{-0.5cm}
\end{figure}
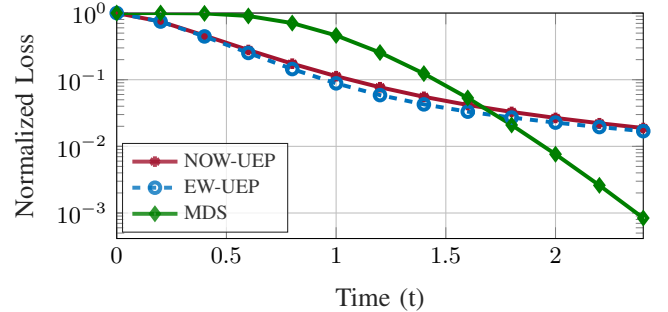

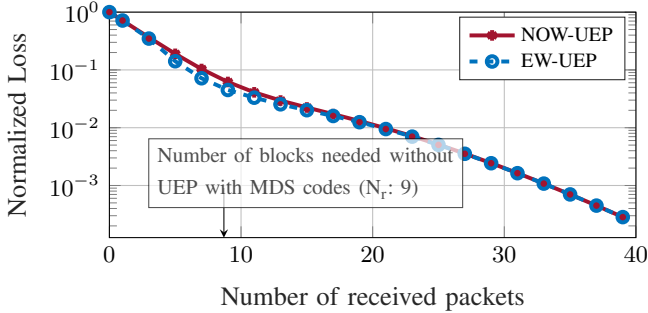
\begin{figure}[t]
	\centering
	\vspace{-3cm}
	\input{tikz_files/three_classes_UEP_MDS_N_40workers}
\vspace{-0.25cm}
	\caption{Normalized loss of the estimator using UEP codes with three classes as a function of number of received packets.}  \label{three_cls_mse_N}
	\vspace{-0.5cm}
\end{figure}

We consider multiplication of two matrices $\Av \in \mathbb{R}^{N U\times M}$ and 
$\Bv \in \mathbb{R}^{M\times P Q}$ with the help of $W=40$ workers whose task completion times are modeled by exponential latency model with parameter $\lambda=0.25$. 
At time $t$, the approximation is performed by only using the worker responses up until time $t$, and the rest are ignored. We select $U = Q = 5$, and $M = 100$.

Firstly, as discussed earlier, we classify each row and column blocks of $\Av$ and $\Bv$ with importance levels \textit{high}, \textit{medium}, and \textit{low}. 
The element of each block is i.i.d. and distributed with $\mathcal{N}(0,10)$, $\mathcal{N}(0,1)$, and $\mathcal{N}(0,0.1)$, for \textit{high}, \textit{medium}, and \textit{low} levels, respectively. 
We assume that both $\Av$ and $\Bv$ have only one instance of row and column from each level, i.e., $N = 3$, $P = 3$. 
We take the multiplication of \textit{high} and \textit{high} blocks as \textit{class one}, \textit{high} and \textit{medium} blocks as \textit{class two}, and the remaining as \textit{class three}. 
With this definition, we have $(k_1, k_2,k_3) = (1,2,6)$ sub-blocks in each class. We select the window selection probabilities for both NOW and EW-UEP strategies as $\Gamma_1 = 0.35, \Gamma_2 = 0.35, \ \Gamma_3 = 0.3$.

The decoding probabilities for each class are obtained through the formulation given in \cite{vukobratovic2012unequal}, as also depicted for the NOW-UEP strategy in Fig. \ref{NOW_prob}. 
In Fig. \ref{three_cls_mse}, these decoding probabilities are used to obtain the normalized expected loss values as a function of time $t$ along with the performance obtained with the MDS codes which are also used in \cite{lee2017speeding} for coded computation.
Until time $t = 1.7$, the UEP protection with both NOW and EW is better than that of MDS, since the UEP coding strategy enables early recovery of important classes with a small number of received packets. 
For instance, at time $t = 1$ the MDS coding approach gives a normalized loss of $0.46$ which is extremely high while the EW-UEP strategy provides loss of $0.088$ which is significantly lower and close to perfect recovery. 
In other words, if we are interested in an earlier recovery of certain important parts, using the UEP coding approach for matrix approximation is highly advantageous. 
After time $t= 1.7$, the MDS code starts to perform better than the others since it can fully recover $\Cv$ after receiving nine packets. 
If we wait long enough, the UEP strategy will also fully recover the desired matrix product.

For further interpretation, we give the normalized loss values of matrix multiplication with MDS and approximate matrix multiplication using NOW and EW-UEP coding in Fig. \ref{three_cls_mse_N} as a function of the number of received packets. 
The matrix multiplication with MDS codes needs to receive $\sum_{l \in [L]} k_l$ packets to fully recover the result, where $k_l$ is the number of packets in the $l$-th level.
Receiving less than $\sum_{l \in [L]} k_l$ will not provide any partial information, and results in no recovery, hence the normalized loss with MDS coding is unity until it receives nine packets (the minimum required for recovery). 
However, matrix approximation with NOW and EW-UEP coding strategies start to recover more important classes after receiving only very few packets, and continue to provide additional partial information after each received block.

It is also worth noting that we choose the window selection distributions for the UEP codes arbitrarily. 
As a further improvement, this distribution can be optimized to minimize the loss in the matrix approximation. 

\subsection{ML Performance with UEP Coded Matrix Multiplication}
\label{subsec: ML performance}
In this section, we investigate the accuracy of several trained DNN models which employ NOW-UEP, EW-UEP, and block repetition coding to calculate the gradient matrix. More specifically, we present the classification results for the MNIST dataset, containing 60,000 training samples with the DNN model defined in Table \ref{tab:model_layers}. The model is trained using Stochastic Gradient Descent (SGD), with a learning rate of 0.01, and cross-entropy as loss. The images are passed through the model in mini batches of 64 at a time over two or three epochs, depending on the learning speed. 
\begin{table}[t]
\vspace{0.5cm}
\footnotesize
    \centering
        \caption{A summary of the model layers of the DNN.}
    \label{tab:model_layers}
\begin{tabular}{|c|c|c|c|}
\hline
Layer       & Weight Dim    & Grad Input Dim & Grad Weight Dim              \\ \hline
Dense 1     & (784x100)   & (64x100)x(100x784)   & (784x64)x(64x100)     \\ \hline
ReLU        &   -           & -                     &   -                       \\ \hline
Dense 2     & (100x200)   & (64x200)x(200x100)   & (100x64)x(64x200)      \\ \hline
ReLU        &   -           & -                     &   -                       \\ \hline
Dense 3     & (200x10)   & (64x10)x(10x200)   & (200x64)x(64x10)      \\ \hline
Softmax     &   -           & -                     &   -                       \\ \hline
\end{tabular}
\end{table}
\begin{table}[t]
\footnotesize
    \centering
        \caption{A summary of the encoding parameters in Section \ref{subsec: ML performance}.}
    \label{tab:model_parameters}
\begin{tabular}{|c|c|c|}
\hline
Encoding Type  & $W$ & $\Omega$ \\ \hline
Uncoded   & 9 & 9 / 9                  \\ \hline
NOW/EW - UEP &  15 & 9 / 15                 \\ \hline
Block Repetition &  18 & 9 / 18                 \\ \hline  
\end{tabular}
\vspace{-0.5cm}
\end{table}
To fairly compare different codes, we scale the time required to complete a task as $F(\Omega t)$, where $\Omega$ is the number of matrix sub-products over the number of workers.
%
In this scaling, we assume that doubling the number of workers halves the average completion time.
Since $N = P = 3$, the number of matrix sub-products required for successful computation is $N P = 9$.
%
For the simulations, we use $\lambda=0.5$ with an exponential latency model and $T_{\max} = 0.25, 0.5, 1, 2$. We also show the ideal scenario in which the matrix multiplications are centralized and there are no stragglers as a benchmark.
We employ the parameters given in Table \ref{tab:model_parameters} for encoding the gradients in each dense layer shown in Table \ref{tab:model_layers}.  To ensure that the higher weight values are biased towards the same portion of the matrix, the indexes are permuted in a descending order of the column/row weights before using the UEP coding. This permutation idea is similar to the one used in \cite{yuster2005fast} which proposes a fast matrix multiplication algorithm. 

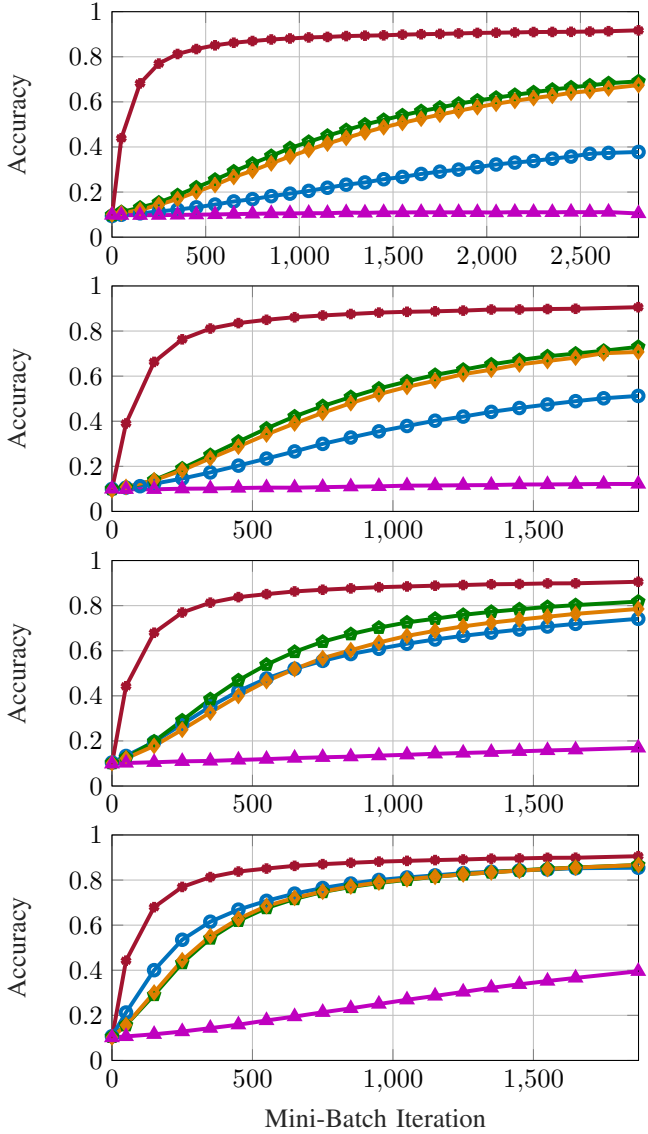
\begin{figure}[t]
\begin{subfigure}[b]{0.3\textwidth}
\vspace{-3.6cm}
    \input{tikz_files/MNIST_Classification_tmax0.25}
\end{subfigure}
\begin{subfigure}[b]{0.3\textwidth}
\vspace{-3.5cm}
    \input{tikz_files/MNIST_Classification_tmax0.5}
\end{subfigure}
\begin{subfigure}[b]{0.3\textwidth}
\vspace{-3.5cm}
	\input{tikz_files/MNIST_Classification_tmax1}
\end{subfigure}
\begin{subfigure}[b]{0.3\textwidth}
\vspace{-3.5cm}
  	\input{tikz_files/MNIST_Classification_tmax2}
\end{subfigure}
	\caption{MNIST classification accuracy for the neural network proposed in Section \ref{sec:Numerical Examples} with $\lambda =  0.5$, and $T_{\max} = 0.25, 0.5, 1$, and $2$, respectively.} \label{MNIST}
	\vspace{-0.6cm}
\end{figure}

We observe from the results shown in Fig. \ref{MNIST} that the UEP coding strategy shows a significant advantage in providing a closer approximation with a shorter wait time, e.g., for $T_{\max} = 0.25$ and $0.5$. Hence, the accuracy of the DNN based learning algorithm is improved. On the other hand, employing block repetitions increases the number of workers required and does not result in a better approximation compared to the uncoded scheme at any point.

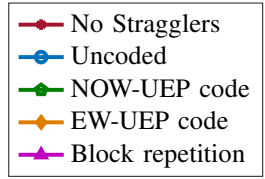
\begin{figure}[t]
\hspace{-0.15cm}
    \input{tikz_files/MNIST_Captions}
    \vspace{-3.3cm}
    \caption{Legends used in Fig. \ref{MNIST}.}
\end{figure}

\section{Conclusions} \label{sec: conclusion}
We have studied distributed approximate matrix multiplication using UEP codes with the objective of mitigating the stragglers and speeding up the large-scale operations which are common in machine learning and data mining algorithms.
We use UEP codes to provide better protection for the sub-operations which have higher effects on the resulting matrix product by better protecting the sub-operations with larger norms.
We validate the effectiveness of the proposed approach through an analytical assessments based on simplified models for sparse matrices, and compare our results with those obtained with MDS codes via simulations. 
Furthermore, the proposed strategy is applied to the backpropagation steps of a DNN.
Our results clearly show that, in the presence of stragglers, we can have a performance closer to the centralized training earlier by striking a balance between the precision of the updates and the computation deadline.

\bibliographystyle{IEEEtran}
\bibliography{bibs_matrix_mult}
\end{document}

%% file: tikz_files/system_model.tex
\begin{tikzpicture}[->,>=stealth',auto,node distance=3cm,
  thick,main node/.style={circle,draw,font=\sffamily\Large\bfseries},scale=0.9, every node/.style={scale=0.74}]
\definecolor{apricot}{rgb}{0.98, 0.81, 0.69}
\definecolor{antiquebrass}{rgb}{0.8, 0.58, 0.46}
\definecolor{arylideyellow}{rgb}{0.91, 0.84, 0.42}
\definecolor{bananamania}{rgb}{0.98, 0.91, 0.71}
\definecolor{babyblue}{rgb}{0.54, 0.81, 0.94}
\definecolor{babypink}{rgb}{0.96, 0.76, 0.76}
\definecolor{caribbeangreen}{rgb}{0.0, 0.8, 0.6}
\definecolor{celadon}{rgb}{0.67, 0.88, 0.69}
\definecolor{cornsilk}{rgb}{1.0, 0.97, 0.86}	
	
\tikzstyle{arrow} = [thick,->,>=stealth]

\node at (2.75, 2){Parameter Server (PS)};
\draw[rounded corners, fill=bananamania] (-0.85,1.6) rectangle (6.25,-3.7);
\node at (1.25, 1.25){$\mathbf{A}$ ($NU \times M$)};
\node at (4.15, 1.25){$\mathbf{B}$ ($M \times PQ$)};
\draw[fill=babyblue] (0,0) rectangle (2.5,0.5);
\node at (-0.5, 0.25){$\mathbf{A}_1$};
\node at (1.25,0.25){};
\draw[fill=babyblue] (0,-0.5) rectangle (2.5,0);
\node at (-0.5, -0.1){$\vdots$};
\node at (1.25,-0.25){};
\draw[fill=babyblue] (0,-1) rectangle (2.5,-0.5);
\node at (-0.5, -0.75){$\mathbf{A}_n$};
\node at (1.25,-0.75){$U \times M$};
\draw[fill=babyblue] (0,-1.5) rectangle (2.5,-1);
\node at (-0.5, -1.1){$\vdots$};
\node at (1.25,-1.25){};
\draw[fill=babyblue] (0,-2) rectangle (2.5,-1.5);
\node at (-0.5, -1.75){$\mathbf{A}_N$};
\node at (1.25,-1.75){};
\node (Ai) at (1.25,-2){};

\draw [fill=babypink](3,0.5) rectangle (3.5,-2);
\node at (3.25, 0.75){$\mathbf{B}_1$};

\draw [fill=babypink](3.5,0.5) rectangle (4,-2);
\node at (3.75, 0.75){$\cdots$};

\draw [fill=babypink](4,0.5) rectangle (4.5,-2);
\node at (4.25, 0.75){$\mathbf{B}_p$};
\node at (4.25,-0.35){$M$};
\node at (4.25,-0.65){$\times$};
\node at (4.25,-0.95){$Q$};

\draw [fill=babypink](4.5,0.5) rectangle (5,-2);
\node at (4.75, 0.75){$\cdots$};

\draw [fill=babypink](5,0.5) rectangle (5.5,-2);
\node at (5.25, 0.75){$\mathbf{B}_P$};
\node (B) at (6.25,-1.6){};
\node (Bj) at (4.25,-2){};
	
\node[draw,rectangle, rounded corners, fill=celadon, font=\small \sffamily] (W1) at (0,-4.85) {Worker $1$};
\node at (1.4, -4.85){$\cdots$};
\node[draw,rectangle, rounded corners, fill=celadon, font=\small \sffamily] (Ww) at (2.78,-4.85) {Worker $w$};
\node at (4.1, -4.85){$\cdots$};
\node[draw,rectangle, rounded corners, fill=celadon,font=\small \sffamily] (W1) at (5.5,-4.85) {Worker $W$};
\node (W) at (6.45,-4.85){};
\node (Wij) at (2.75,-3.25){};
\node[draw,rectangle, rounded corners, fill=cornsilk, font=\small \sffamily] (WA) at (2.1,-3.3) {Enc A.};
\node[draw,rectangle, rounded corners, fill=cornsilk, font=\small \sffamily] (WB) at (3.5,-3.3) {Enc B.};

\node (e_out) at (2.75,-3.5){};

\draw [->] (W) to [out = 60, in=-45] node[anchor=west] {$\mathbf{W}_A^w \mathbf{W}_B^w$} (B);

\draw [->] (Ai) to  (WA);
\node at (1.45,-2.65){$\mathbf{A}_n$};
\draw [->] (Bj) to  (WB);
\node at (4.25,-2.65){$\mathbf{B}_p$};
\draw [->] (WA) to node[anchor=east] {$\mathbf{W}_A^w$} (Ww);
\draw [->] (WB) to node[anchor=west] {$\mathbf{W}_B^w$} (Ww);

 \end{tikzpicture}

%% file: tikz_files/NOW_prob_40workers.tex
\begin{tikzpicture}
\definecolor{mycolor1}{rgb}{0.63529,0.07843,0.18431}%
\definecolor{mycolor2}{rgb}{0.00000,0.44706,0.74118}%
\definecolor{mycolor3}{rgb}{0.00000,0.49804,0.00000}%
\definecolor{mycolor4}{rgb}{0.87059,0.49020,0.00000}%
\definecolor{mycolor5}{rgb}{0.00000,0.44700,0.74100}%
\definecolor{mycolor6}{rgb}{0.74902,0.00000,0.74902}%

\begin{axis}[%
font=\small,
width=7cm,
height=3cm,
scale only axis,
xmin=0,
xmax=40,
xlabel style={font=\color{white!15!black}},
xlabel={Received packets (N)},
ymin=0,
ymax=1,
ylabel style={font=\color{white!15!black}},
ylabel={Decoding probabilities},
axis background/.style={fill=white},
title style={font=\bfseries},
xmajorgrids,
ymajorgrids,
legend style={legend cell align=left, align=left, draw=white!15!black, nodes={scale=0.85, transform shape}, at={(0.7,0.25)}, anchor=west, fill opacity=0.8}
]
\addplot [color=mycolor1, line width=1.5pt, mark=o, mark options={solid, mycolor1}]
  table[row sep=crcr]{%
0	0\\
2	0.5775\\
4	0.82149375\\
6	0.924581109375\\
8	0.968135518710937\\
10	0.98653725665537\\
12	0.994311990936893\\
14	0.997596816170837\\
16	0.998984654832178\\
18	0.999571016666595\\
20	0.999818754541636\\
22	0.99992342379384\\
24	0.999967646552897\\
26	0.999986330668598\\
28	0.999994224707481\\
30	0.99999755993891\\
32	0.999998969074188\\
34	0.999999564433844\\
36	0.999999815973297\\
38	0.999999922248718\\
40	0.999999967150082\\
};
\addlegendentry{$P_{d,1}(N)$}

\addplot [color=mycolor2, line width=1.5pt, mark=x, mark options={solid, mycolor2}]
  table[row sep=crcr]{%
0	0\\
2	0.1225\\
4	0.43701875\\
6	0.680920078125\\
8	0.830873137773437\\
10	0.914045561722753\\
12	0.957558701606056\\
14	0.979480507304845\\
16	0.990237065694028\\
18	0.995413178204368\\
20	0.997866880374646\\
22	0.999016290274732\\
24	0.999549540467271\\
26	0.99979496002899\\
28	0.999907151066459\\
30	0.999958143567483\\
32	0.999981205429465\\
34	0.999991590222704\\
36	0.999996248686491\\
38	0.999998331337894\\
40	0.999999259613426\\
};
\addlegendentry{$P_{d,2}(N)$}

\addplot [color=mycolor3, line width=1.5pt, mark=diamond, mark options={solid, mycolor3}]
  table[row sep=crcr]{%
0	0\\
2	0\\
4	0\\
6	0.000729\\
8	0.01129221\\
10	0.0473489874\\
12	0.117848738862\\
14	0.21948416523063\\
16	0.340217674088892\\
18	0.465619902225398\\
20	0.583629170552518\\
22	0.686587214036805\\
24	0.771191606986015\\
26	0.837416514972112\\
28	0.887210956393012\\
30	0.923405247992448\\
32	0.948965004355415\\
34	0.966570375603619\\
36	0.978435996647857\\
38	0.986281620605649\\
40	0.991381946688624\\
};
\addlegendentry{$P_{d,3}(N)$}

\end{axis}

\begin{axis}[%
width=6.771in,
height=2.604in,
at={(0in,0in)},
scale only axis,
xmin=0,
xmax=1,
ymin=0,
ymax=1,
axis line style={draw=none},
ticks=none,
axis x line*=bottom,
axis y line*=left
]
\end{axis}
\end{tikzpicture}%

%% file: tikz_files/three_classes_UEP_t_40workers.tex
\begin{tikzpicture}
\definecolor{mycolor1}{rgb}{0.63529,0.07843,0.18431}%
\definecolor{mycolor2}{rgb}{0.00000,0.44706,0.74118}%
\definecolor{mycolor3}{rgb}{0.00000,0.49804,0.00000}%
\definecolor{mycolor4}{rgb}{0.87059,0.49020,0.00000}%
\definecolor{mycolor5}{rgb}{0.00000,0.44700,0.74100}%
\definecolor{mycolor6}{rgb}{0.74902,0.00000,0.74902}%

\begin{semilogyaxis}[%
font=\small,
width=7cm,
height=3cm,
scale only axis,
xmin=0,
xmax=2.4,
xlabel style={font=\color{white!15!black}},
xlabel={Time (t)},
ymax=1,
ylabel style={font=\color{white!15!black}},
ylabel={Normalized Loss},
axis background/.style={fill=white},
xmajorgrids,
ymajorgrids,
legend style={legend cell align=left, align=left, draw=white!15!black, nodes={scale=0.85, transform shape}, at={(0.01,0.225)}, anchor=west, fill opacity=0.8}
]

\addplot [color=mycolor1, line width=1.5pt, mark=asterisk, mark options={solid, mycolor1}]
  table[row sep=crcr]{%
0	1\\
0.2	0.745408499508458\\
0.4	0.459511336721325\\
0.6	0.279078378482209\\
0.8	0.173910594878883\\
1	0.112957558447755\\
1.2	0.077077277280897\\
1.4	0.0553823255567802\\
1.6	0.0417777983249171\\
1.8	0.0328550230477046\\
2	0.0267007560129081\\
2.2	0.0222340156757181\\
2.4	0.0188376281931689\\
2.6	0.0161534744325889\\
};
\addlegendentry{NOW-UEP}

\addplot [color=mycolor2, dashed, line width=1.5pt, mark=o, mark options={solid, mycolor2}]
  table[row sep=crcr]{%
0	1\\
0.2	0.74561015651208\\
0.4	0.445979499823152\\
0.6	0.252123028287349\\
0.8	0.144483635475889\\
1	0.0881713704027316\\
1.2	0.0587324625454995\\
1.4	0.0426422262709409\\
1.6	0.0331550269972291\\
1.8	0.0270366357380174\\
2	0.022736677778323\\
2.2	0.0194921144252786\\
2.4	0.0169106928821743\\
2.6	0.0147798990355951\\
};
\addlegendentry{EW-UEP}

\addplot [color=mycolor3, line width=1.5pt, mark=diamond, mark options={solid, mycolor3}]
  table[row sep=crcr]{%
0	1\\
0.2	0.99989270726923\\
0.4	0.98856807850022\\
0.6	0.904377279645571\\
0.8	0.706927850184667\\
1	0.461470058007696\\
1.2	0.255686212970652\\
1.4	0.123452026419275\\
1.6	0.0532362122164085\\
1.8	0.0209226336470409\\
2	0.00761511288434526\\
2.2	0.00259928192533236\\
2.4	0.000840368283381599\\
2.6	0.00025940379572753\\
};
\addlegendentry{MDS}

\end{semilogyaxis}
\end{tikzpicture}%

%% file: tikz_files/three_classes_UEP_MDS_N_40workers.tex
\begin{tikzpicture}
\definecolor{mycolor1}{rgb}{0.63529,0.07843,0.18431}%
\definecolor{mycolor2}{rgb}{0.00000,0.44706,0.74118}%
\definecolor{mycolor3}{rgb}{0.00000,0.49804,0.00000}%
\definecolor{mycolor4}{rgb}{0.87059,0.49020,0.00000}%
\definecolor{mycolor5}{rgb}{0.00000,0.44700,0.74100}%
\definecolor{mycolor6}{rgb}{0.74902,0.00000,0.74902}%

\begin{semilogyaxis}[%
font=\small,
width=7cm,
height=3cm,
scale only axis,
xmin=0,
xmax=40,
xlabel style={font=\color{white!15!black}},
xlabel={Number of received packets},
ymax=1,
ylabel style={font=\color{white!15!black}},
ylabel={Normalized Loss},
axis background/.style={fill=white},
xmajorgrids,
ymajorgrids,
legend style={nodes={scale=0.85, transform shape}, legend cell align=left, align=left, draw=white!15!black}
]
\addplot [color=mycolor1, line width=1.5pt, mark=asterisk, mark options={solid, mycolor1}]
  table[row sep=crcr]{%
0	1\\
1	0.715932148364581\\
3	0.365534453372291\\
5	0.189767115088061\\
7	0.103693130763534\\
9	0.0618593898136343\\
11	0.0409514133547276\\
13	0.0295458780342457\\
15	0.0223698357584026\\
17	0.0171724011993714\\
19	0.0130802222730827\\
21	0.00978074403830491\\
23	0.00715039924499832\\
25	0.00510740460368202\\
27	0.00356760902428653\\
29	0.00244080784776147\\
31	0.00163840139590627\\
33	0.00108087600239665\\
35	0.00070191994458443\\
37	0.000449343390482038\\
39	0.000283926181783713\\
};
\addlegendentry{NOW-UEP}

\addplot [color=mycolor2, dashed, line width=1.5pt, mark=o, mark options={solid, mycolor2}]
  table[row sep=crcr]{%
0	1\\
1	0.715932148364581\\
3	0.348632416199984\\
5	0.141285609934259\\
7	0.0717197242005524\\
9	0.0449644688804384\\
11	0.0330769939238417\\
13	0.0253995483270316\\
15	0.0200871638090778\\
17	0.015950899590854\\
19	0.0124591576586648\\
21	0.00947929236871762\\
23	0.00700902035945513\\
25	0.00504260820640437\\
27	0.00353834375825671\\
29	0.00242771075911545\\
31	0.00163257414494092\\
33	0.00107829344064254\\
35	0.000700778641054915\\
37	0.000448840148240035\\
39	0.000283704702679543\\
};
\addlegendentry{EW-UEP}

\end{semilogyaxis}

\begin{axis}[%
width=6.771in,
height=2.604in,
at={(0in,0in)},
scale only axis,
xmin=0,
xmax=1,
ymin=0,
ymax=1,
axis line style={draw=none},
ticks=none,
axis x line*=bottom,
axis y line*=left
]
\draw[-{stealth}, color=black] (axis cs:0.0885,0.10) -- (axis cs:0.0885,0.0);
\node[below right, align=left, draw=black,fill= white, opacity=0.7]
at (rel axis cs:0.03,0.2) {\footnotesize Number of blocks needed without\\\footnotesize UEP with MDS codes $\text{(N}_\text{r}\text{: 9)}$};
\end{axis}
\end{tikzpicture}%


%% file: tikz_files/MNIST_Captions.tex
\begin{tikzpicture}
\definecolor{mycolor1}{rgb}{0.63529,0.07843,0.18431}%
\definecolor{mycolor2}{rgb}{0.00000,0.44706,0.74118}%
\definecolor{mycolor3}{rgb}{0.00000,0.49804,0.00000}%
\definecolor{mycolor4}{rgb}{0.87059,0.49020,0.00000}%
\definecolor{mycolor5}{rgb}{0.00000,0.44700,0.74100}%
\definecolor{mycolor6}{rgb}{0.74902,0.00000,0.74902}%

\begin{axis}[%
hide axis,
xmin=0,
xmax=0.4,
ymin=0,
ymax=0.4,
legend style={legend cell align=left, align=left, draw=white!15!black}
]

\addlegendimage{color=mycolor1, line width=1.5pt, mark=asterisk, mark options={solid, mycolor1}}
\addlegendentry{No Stragglers}

\addlegendimage{color=mycolor2, line width=1.5pt, mark=o, mark options={solid, mycolor2}}
\addlegendentry{Uncoded}

\addlegendimage{color=mycolor3, line width=1.5pt, mark=pentagon, mark options={solid, mycolor3}}
\addlegendentry{NOW-UEP code}

\addlegendimage{color=mycolor4, line width=1.5pt, mark=diamond, mark options={solid, mycolor4}}
\addlegendentry{EW-UEP code}

\addlegendimage{color=mycolor6, line width=1.5pt, mark=triangle, mark options={solid, mycolor6}}
\addlegendentry{Block repetition}

\end{axis}

\end{tikzpicture}%

%% file: tikz_files/MNIST_Classification_tmax0.25.tex
\begin{tikzpicture}
\definecolor{mycolor1}{rgb}{0.63529,0.07843,0.18431}%
\definecolor{mycolor2}{rgb}{0.00000,0.44706,0.74118}%
\definecolor{mycolor3}{rgb}{0.00000,0.49804,0.00000}%
\definecolor{mycolor4}{rgb}{0.87059,0.49020,0.00000}%
\definecolor{mycolor5}{rgb}{0.00000,0.44700,0.74100}%
\definecolor{mycolor6}{rgb}{0.74902,0.00000,0.74902}%

\begin{axis}[%
width=7cm,
height=3cm,
scale only axis,
xmin=0,
xmax=2810,
xlabel style={font=\color{white!15!black}},
ymin=0,
ymax=1,
ylabel style={font=\color{white!15!black}},
ylabel={Accuracy},
axis background/.style={fill=white},
xmajorgrids,
ymajorgrids,
legend style={legend cell align=left, align=left, draw=white!15!black, nodes={scale=0.85, transform shape}, at={(0.01,0.65)}, anchor=west, fill opacity=0.8}
]

\addplot [color=mycolor1, line width=1.5pt, mark=asterisk, mark options={solid, mycolor1}]
  table[row sep=crcr]{%
0	0.0945312	\\
50	0.438741	\\
150	0.682084	\\
250	0.769869	\\
350	0.8123	\\
450	0.834178	\\
550	0.850897	\\
650	0.862234	\\
750	0.869825	\\
850	0.876706	\\
950	0.881141	\\
1050	0.885766	\\
1150	0.888022	\\
1250	0.891791	\\
1350	0.893613	\\
1450	0.895028	\\
1550	0.898175	\\
1650	0.900025	\\
1750	0.901506	\\
1850	0.903506	\\
1950	0.90575	\\
2050	0.9068	\\
2150	0.907759	\\
2250	0.909672	\\
2350	0.910122	\\
2450	0.910675	\\
2550	0.911872	\\
2650	0.913391	\\
2810	0.917188	\\
};

\addplot [color=mycolor2, line width=1.5pt, mark=o, mark options={solid, mycolor2}]
  table[row sep=crcr]{%
0	0.09375	\\
50	0.0993125	\\
150	0.105644	\\
250	0.113812	\\
350	0.122075	\\
450	0.132187	\\
550	0.144506	\\
650	0.158362	\\
750	0.168587	\\
850	0.181819	\\
950	0.193081	\\
1050	0.205206	\\
1150	0.219138	\\
1250	0.232706	\\
1350	0.242613	\\
1450	0.25685	\\
1550	0.267006	\\
1650	0.280313	\\
1750	0.290575	\\
1850	0.299525	\\
1950	0.310588	\\
2050	0.322288	\\
2150	0.330769	\\
2250	0.33795	\\
2350	0.347925	\\
2450	0.357269	\\
2550	0.369762	\\
2650	0.373613	\\
2810	0.377812	\\
};

\addplot [color=mycolor3, line width=1.5pt, mark=pentagon, mark options={solid, mycolor3}]
  table[row sep=crcr]{%
0	0.098125	\\
50	0.111238	\\
150	0.131025	\\
250	0.153456	\\
350	0.18605	\\
450	0.220513	\\
550	0.254738	\\
650	0.29145	\\
750	0.326225	\\
850	0.360013	\\
950	0.394563	\\
1050	0.422731	\\
1150	0.44975	\\
1250	0.474775	\\
1350	0.499531	\\
1450	0.519888	\\
1550	0.540219	\\
1650	0.558819	\\
1750	0.574525	\\
1850	0.591719	\\
1950	0.605569	\\
2050	0.618237	\\
2150	0.630794	\\
2250	0.642419	\\
2350	0.654119	\\
2450	0.664169	\\
2550	0.672844	\\
2650	0.682581	\\
2810	0.690937	\\
};

\addplot [color=mycolor4, line width=1.5pt, mark=diamond, mark options={solid, mycolor4}]
  table[row sep=crcr]{%
0	0.100312	\\
50	0.106306	\\
150	0.120619	\\
250	0.143794	\\
350	0.169137	\\
450	0.200712	\\
550	0.231225	\\
650	0.266881	\\
750	0.294513	\\
850	0.327269	\\
950	0.35785	\\
1050	0.386838	\\
1150	0.415	\\
1250	0.440094	\\
1350	0.464481	\\
1450	0.487456	\\
1550	0.506837	\\
1650	0.525544	\\
1750	0.543356	\\
1850	0.560619	\\
1950	0.576438	\\
2050	0.5916	\\
2150	0.604431	\\
2250	0.616869	\\
2350	0.627163	\\
2450	0.640162	\\
2550	0.648538	\\
2650	0.659637	\\
2810	0.674375	\\
};

\addplot [color=mycolor6, line width=1.5pt, mark=triangle, mark options={solid, mycolor6}]
  table[row sep=crcr]{%
0	0.0984375	\\
50	0.09765	\\
150	0.097325	\\
250	0.0975688	\\
350	0.0985938	\\
450	0.0997813	\\
550	0.100769	\\
650	0.102306	\\
750	0.103444	\\
850	0.104888	\\
950	0.105181	\\
1050	0.106275	\\
1150	0.106156	\\
1250	0.10905	\\
1350	0.107731	\\
1450	0.109562	\\
1550	0.109556	\\
1650	0.110181	\\
1750	0.110025	\\
1850	0.109969	\\
1950	0.109963	\\
2050	0.109106	\\
2150	0.1109	\\
2250	0.110425	\\
2350	0.110031	\\
2450	0.1113	\\
2550	0.110337	\\
2650	0.111206	\\
2810	0.105	\\
};

\end{axis}

\begin{axis}[%
width=6.771in,
height=2.604in,
at={(0in,0in)},
scale only axis,
xmin=0,
xmax=1,
ymin=0,
ymax=1,
axis line style={draw=none},
ticks=none,
axis x line*=bottom,
axis y line*=left
]

\end{axis}
\end{tikzpicture}%

%% file: tikz_files/MNIST_Classification_tmax0.5.tex
\begin{tikzpicture}
\definecolor{mycolor1}{rgb}{0.63529,0.07843,0.18431}%
\definecolor{mycolor2}{rgb}{0.00000,0.44706,0.74118}%
\definecolor{mycolor3}{rgb}{0.00000,0.49804,0.00000}%
\definecolor{mycolor4}{rgb}{0.87059,0.49020,0.00000}%
\definecolor{mycolor5}{rgb}{0.00000,0.44700,0.74100}%
\definecolor{mycolor6}{rgb}{0.74902,0.00000,0.74902}%

\begin{axis}[%
width=7cm,
height=3cm,
scale only axis,
xmin=0,
xmax=1874,
xlabel style={font=\color{white!15!black}},
ymin=0,
ymax=1,
ylabel style={font=\color{white!15!black}},
ylabel={Accuracy},
axis background/.style={fill=white},
xmajorgrids,
ymajorgrids,
legend style={legend cell align=left, align=left, draw=white!15!black, nodes={scale=0.85, transform shape}, at={(0.01,0.65)}, anchor=west, fill opacity=0.8}
]

\addplot [color=mycolor1, line width=1.5pt, mark=asterisk, mark options={solid, mycolor1}]
  table[row sep=crcr]{%
0	0.0982813	\\
50	0.3902	\\
150	0.663069	\\
250	0.763344	\\
350	0.811144	\\
450	0.835206	\\
550	0.849844	\\
650	0.861294	\\
750	0.869094	\\
850	0.875575	\\
950	0.881806	\\
1050	0.885325	\\
1150	0.887494	\\
1250	0.8907	\\
1350	0.895444	\\
1450	0.895675	\\
1550	0.897606	\\
1650	0.898812	\\
1873    0.905781 \\
};

\addplot [color=mycolor2, line width=1.5pt, mark=o, mark options={solid, mycolor2}]
  table[row sep=crcr]{%
0	0.0995313	\\
50	0.105619	\\
100	0.111906	\\
150	0.122819	\\
250	0.145663	\\
350	0.173006	\\
450	0.203187	\\
550	0.234444	\\
650	0.266106	\\
750	0.299575	\\
850	0.327769	\\
950	0.355081	\\
1050	0.378931	\\
1150	0.402313	\\
1250	0.42055	\\
1350	0.441487	\\
1450	0.4587	\\
1550	0.474419	\\
1650	0.488963	\\
1750	0.501794	\\
1873    0.512344 \\
};

\addplot [color=mycolor3, line width=1.5pt, mark=pentagon, mark options={solid, mycolor3}]
  table[row sep=crcr]{%
0	0.0995313	\\
50	0.105356	\\
150	0.139237	\\
250	0.190619	\\
350	0.250063	\\
450	0.3095	\\
550	0.367725	\\
650	0.421262	\\
750	0.467194	\\
850	0.507269	\\
950	0.545	\\
1050	0.576219	\\
1150	0.605375	\\
1250	0.628425	\\
1350	0.651881	\\
1450	0.671031	\\
1550	0.687781	\\
1650	0.700069	\\
1750	0.712456	\\
1873    0.729844  \\
};

\addplot [color=mycolor4, line width=1.5pt, mark=diamond, mark options={solid, mycolor4}]
  table[row sep=crcr]{%
0	0.0945312	\\
50	0.108394	\\
150	0.138112	\\
250	0.182225	\\
350	0.235713	\\
450	0.287525	\\
550	0.341325	\\
650	0.390294	\\
750	0.437375	\\
850	0.480312	\\
950	0.519869	\\
1050	0.552275	\\
1150	0.579488	\\
1250	0.608781	\\
1350	0.628625	\\
1450	0.651894	\\
1550	0.667169	\\
1650	0.682119	\\
1750	0.7016	\\
1873    0.707344\\
};

\addplot [color=mycolor6, line width=1.5pt, mark=triangle, mark options={solid, mycolor6}]
  table[row sep=crcr]{%
0	0.1	\\
50	0.0981875	\\
150	0.09815	\\
250	0.10075	\\
350	0.101462	\\
450	0.103813	\\
550	0.105569	\\
650	0.105256	\\
750	0.107519	\\
850	0.109744	\\
950	0.111212	\\
1050	0.114356	\\
1150	0.11485	\\
1250	0.116194	\\
1350	0.116937	\\
1450	0.11945	\\
1550	0.119706	\\
1650	0.120675	\\
1750	0.121794	\\
1873 0.121719 \\
};

\end{axis}

\begin{axis}[%
width=6.771in,
height=2.604in,
at={(0in,0in)},
scale only axis,
xmin=0,
xmax=1,
ymin=0,
ymax=1,
axis line style={draw=none},
ticks=none,
axis x line*=bottom,
axis y line*=left
]

\end{axis}
\end{tikzpicture}%

%% file: tikz_files/MNIST_Classification_tmax1.tex
\begin{tikzpicture}
\definecolor{mycolor1}{rgb}{0.63529,0.07843,0.18431}%
\definecolor{mycolor2}{rgb}{0.00000,0.44706,0.74118}%
\definecolor{mycolor3}{rgb}{0.00000,0.49804,0.00000}%
\definecolor{mycolor4}{rgb}{0.87059,0.49020,0.00000}%
\definecolor{mycolor5}{rgb}{0.00000,0.44700,0.74100}%
\definecolor{mycolor6}{rgb}{0.74902,0.00000,0.74902}%

\begin{axis}[%
width=7cm,
height=3cm,
scale only axis,
xmin=0,
xmax=1874,
xlabel style={font=\color{white!15!black}},
ymin=0,
ymax=1,
ylabel style={font=\color{white!15!black}},
ylabel={Accuracy},
axis background/.style={fill=white},
xmajorgrids,
ymajorgrids,
]

\addplot [color=mycolor1, line width=1.5pt, mark=asterisk, mark options={solid, mycolor1}]
  table[row sep=crcr]{%
0	0.0982813	\\
50	0.442359	\\
150	0.679747	\\
250	0.769481	\\
350	0.813197	\\
450	0.837725	\\
550	0.850991	\\
650	0.863003	\\
750	0.8705	\\
850	0.876469	\\
950	0.881769	\\
1050	0.885069	\\
1150	0.888487	\\
1250	0.891259	\\
1350	0.894841	\\
1450	0.895941	\\
1550	0.898622	\\
1650	0.899031	\\
1873	0.905781	\\
};

\addplot [color=mycolor2, line width=1.5pt, mark=o, mark options={solid, mycolor2}]
  table[row sep=crcr]{%
0	0.104375	\\
50	0.133247	\\
150	0.195278	\\
250	0.27425	\\
350	0.352156	\\
450	0.421072	\\
550	0.476794	\\
650	0.519975	\\
750	0.55505	\\
850	0.585381	\\
950	0.608606	\\
1050	0.631075	\\
1150	0.649906	\\
1250	0.666081	\\
1350	0.680363	\\
1450	0.693928	\\
1550	0.707353	\\
1650	0.719178	\\
1873	0.741563	\\
};

\addplot [color=mycolor3, line width=1.5pt, mark=pentagon, mark options={solid, mycolor3}]
  table[row sep=crcr]{%
0	0.102031	\\
50	0.12495	\\
150	0.197728	\\
250	0.291066	\\
350	0.384984	\\
450	0.468731	\\
550	0.538344	\\
650	0.595131	\\
750	0.640009	\\
850	0.673691	\\
950	0.702156	\\
1050	0.725581	\\
1150	0.742003	\\
1250	0.759059	\\
1350	0.772534	\\
1450	0.783788	\\
1550	0.79415	\\
1650	0.802194	\\
1873	0.817969	\\
};

\addplot [color=mycolor4, line width=1.5pt, mark=diamond, mark options={solid, mycolor4}]
  table[row sep=crcr]{%
0	0.101719	\\
50	0.122466	\\
150	0.177769	\\
250	0.249897	\\
350	0.325938	\\
450	0.399625	\\
550	0.465328	\\
650	0.518741	\\
750	0.566956	\\
850	0.602363	\\
950	0.636556	\\
1050	0.665719	\\
1150	0.688234	\\
1250	0.708003	\\
1350	0.724434	\\
1450	0.738241	\\
1550	0.750681	\\
1650	0.763481	\\
1873	0.785312	\\
};

\addplot [color=mycolor6, line width=1.5pt, mark=triangle, mark options={solid, mycolor6}]
  table[row sep=crcr]{%
0	0.0984375	\\
50	0.102225	\\
150	0.105328	\\
250	0.109522	\\
350	0.111253	\\
450	0.115688	\\
550	0.119297	\\
650	0.123906	\\
750	0.127387	\\
850	0.130216	\\
950	0.134722	\\
1050	0.138628	\\
1150	0.142941	\\
1250	0.146563	\\
1350	0.149784	\\
1450	0.153906	\\
1550	0.157753	\\
1650	0.1609	\\
1873	0.169219	\\
};

\end{axis}

\begin{axis}[%
width=6.771in,
height=2.604in,
at={(0in,0in)},
scale only axis,
xmin=0,
xmax=1,
ymin=0,
ymax=1,
axis line style={draw=none},
ticks=none,
axis x line*=bottom,
axis y line*=left
]

\end{axis}
\end{tikzpicture}%

%% file: tikz_files/MNIST_Classification_tmax2.tex
\begin{tikzpicture}
\definecolor{mycolor1}{rgb}{0.63529,0.07843,0.18431}%
\definecolor{mycolor2}{rgb}{0.00000,0.44706,0.74118}%
\definecolor{mycolor3}{rgb}{0.00000,0.49804,0.00000}%
\definecolor{mycolor4}{rgb}{0.87059,0.49020,0.00000}%
\definecolor{mycolor5}{rgb}{0.00000,0.44700,0.74100}%
\definecolor{mycolor6}{rgb}{0.74902,0.00000,0.74902}%

\begin{axis}[%
width=7cm,
height=3cm,
scale only axis,
xmin=0,
xmax=1874,
xlabel style={font=\color{white!15!black}},
xlabel={Mini-Batch Iteration},
ymin=0,
ymax=1,
ylabel style={font=\color{white!15!black}},
ylabel={Accuracy},
axis background/.style={fill=white},
xmajorgrids,
ymajorgrids,
legend style={legend cell align=left, align=left, draw=white!15!black, nodes={scale=0.85, transform shape}, at={(0.01,0.65)}, anchor=west, fill opacity=0.8}
]

\addplot [color=mycolor1, line width=1.5pt, mark=asterisk, mark options={solid, mycolor1}]
  table[row sep=crcr]{%
0	0.0982813	\\
50	0.442359	\\
150	0.679747	\\
250	0.769481	\\
350	0.813197	\\
450	0.837725	\\
550	0.850991	\\
650	0.863003	\\
750	0.8705	\\
850	0.876469	\\
950	0.881769	\\
1050	0.885069	\\
1150	0.888487	\\
1250	0.891259	\\
1350	0.894841	\\
1450	0.895941	\\
1550	0.898622	\\
1650	0.899031	\\
1873	0.905781	\\
};

\addplot [color=mycolor2, line width=1.5pt, mark=o, mark options={solid, mycolor2}]
  table[row sep=crcr]{%
0	0.107344	\\
50	0.212541	\\
150	0.400337	\\
250	0.535244	\\
350	0.615272	\\
450	0.668722	\\
550	0.707788	\\
650	0.739838	\\
750	0.764691	\\
850	0.784525	\\
950	0.80035	\\
1050	0.811434	\\
1150	0.820756	\\
1250	0.830628	\\
1350	0.836956	\\
1450	0.842484	\\
1550	0.847081	\\
1650	0.852525	\\
1873	0.854688	\\
};

\addplot [color=mycolor3, line width=1.5pt, mark=pentagon, mark options={solid, mycolor3}]
  table[row sep=crcr]{%
0	0.105156	\\
50	0.156437	\\
150	0.290778	\\
250	0.432197	\\
350	0.541278	\\
450	0.619691	\\
550	0.676266	\\
650	0.716306	\\
750	0.747322	\\
850	0.769422	\\
950	0.786781	\\
1050	0.801278	\\
1150	0.814528	\\
1250	0.825534	\\
1350	0.834287	\\
1450	0.842872	\\
1550	0.849988	\\
1650	0.8556	\\
1873	0.866719	\\
};

\addplot [color=mycolor4, line width=1.5pt, mark=diamond, mark options={solid, mycolor4}]
  table[row sep=crcr]{%
0	0.1025	\\
50	0.15685	\\
150	0.300897	\\
250	0.444053	\\
350	0.551659	\\
450	0.627916	\\
550	0.684997	\\
650	0.721278	\\
750	0.748581	\\
850	0.771772	\\
950	0.789225	\\
1050	0.803466	\\
1150	0.814669	\\
1250	0.825478	\\
1350	0.834213	\\
1450	0.841512	\\
1550	0.849113	\\
1650	0.855047	\\
1874	0.866719	\\
};

\addplot [color=mycolor6, line width=1.5pt, mark=triangle, mark options={solid, mycolor6}]
  table[row sep=crcr]{%
0	0.101406	\\
50	0.106528	\\
150	0.116025	\\
250	0.128109	\\
350	0.143194	\\
450	0.158459	\\
550	0.177263	\\
650	0.194612	\\
750	0.213588	\\
850	0.230875	\\
950	0.25015	\\
1050	0.269078	\\
1150	0.28605	\\
1250	0.303831	\\
1350	0.322091	\\
1450	0.337603	\\
1550	0.352172	\\
1650	0.365991	\\
1873	0.395469	\\
};

\end{axis}

\begin{axis}[%
width=6.771in,
height=2.604in,
at={(0in,0in)},
scale only axis,
xmin=0,
xmax=1,
ymin=0,
ymax=1,
axis line style={draw=none},
ticks=none,
axis x line*=bottom,
axis y line*=left
]

\end{axis}
\end{tikzpicture}%